\documentclass{article}
\usepackage[dvipdfmx]{graphicx}
\usepackage[numbers, sort, compress]{natbib}
\usepackage{amsmath,amssymb}
\usepackage{bm}

\begin{document}
\title{Evolution of Activity-Dependent Adaptive Boolean Networks towards Criticality: An Analytic Approach}

\author{Taichi Haruna$\ ^{\rm 1}$ \\
\footnotesize{$\ ^{\rm 1}$ Department of Information and Sciences, Tokyo Woman's Christian University} \\
\footnotesize{2-6-1 Zempukuji, Suginami-ku, Tokyo 167-8585, Japan} \\
\footnotesize{E-mail: tharuna@lab.twcu.ac.jp}
}

\date{}
\maketitle

\begin{abstract}
We propose new activity-dependent adaptive Boolean networks inspired by the \textit{cis}-regulatory mechanism in gene regulatory networks. We analytically show that our model can be solved for stationary in-degree distribution for a wide class of update rules by employing the annealed approximation of Boolean network dynamics and that evolved Boolean networks have a preassigned average sensitivity that can be set independently of update rules if certain conditions are satisfied. In particular, when it is set to $1$, our theory predicts that the proposed network rewiring algorithm drives Boolean networks towards criticality. We verify that these analytic results agree well with numerical simulations for four representative update rules. We also discuss the relationship between sensitivity of update rules and stationary in-degree distributions and compare it with that in real-world gene regulatory networks. 
\end{abstract}

\section{Introduction}
\label{sec:intro}
Boolean networks (BNs) \citep{Drossel2008} were originally proposed as a model of gene regulatory networks (GRNs) by S. Kauffman in 1969 \citep{Kauffman1969}. Since then, it also has been used as a useful representation for modeling other complex systems such as neuronal networks \citep{Kurten1988} and social networks \citep{Paczuski2000}. Although Boolean abstraction of real-world complex systems ignores fine details of them, it enables us to study some important aspects of their generic features. One such feature of BNs is the phase transition between ordered phase and disordered phase \citep{Derrida1986a}. It has often been argued, but still is controversial, that real-world living systems such as GRNs and neuronal networks adjust their dynamical behavior towards the boundary between the two phases, criticality \citep{Beggs2003,Petermann2009,Balleza2008,Nykter2008,Valverde2015}. The advantages of criticality also have been studied: Optimal computational ability \citep{Bertschinger2004,Goudarzi2012}, maximal sensitivity to external stimuli \citep{Kinouchi2006}, maximal memory capacity \citep{Haldeman2005} and so on. 

So far, many plausible theoretical models of network self-organization towards criticality have been proposed although the exact mechanisms in gene regulatory or neuronal networks have not yet been known. For example, the following literatures discuss biologically inspired mechanisms: Hebbian learning \citep{Bornholdt2003,Rybarsch2014}, spike-timing dependent plasticity \citep{Meisel2009,Rubinov2011}, dynamical synapses \citep{Levina2007,Levina2009} and homeostatic plasticity \citep{Droste2013} for neuronal networks and local control of feedback loops \citep{MacArthur2010} and adaptation towards both adaptability and stability \citep{Lee2014} for gene regulatory networks. Such models have been collectively called adaptive networks, in which network structure and network state coevolve, and have been paid much attention recently \citep{Gross2009, Sayama2013}. The study of adaptive networks originates from the work by Bornholdt and Rohlf \citep{Bornholdt2000}, which is also motivated by a preliminary work on the relationship between network structure and network state \citep{Christensen1998}. They showed that a simple activity-dependent rewiring rule based on measurement of local dynamics drives random threshold networks towards criticality by numerical simulation. The model has been extended to different situations: Liu and Bassler \citep{Liu2006} reported that the activity-dependent rewiring rule drives random Boolean networks towards criticality by numerical simulation. Recently, this model was extended to networks with modular structure \citep{Gorski2016}. In spite of the structural constraint, self-organization towards criticality was shown to be preserved. Rohlf introduced an activity-dependent threshold change into the original Bornholdt-Rohlf model \citep{Rohlf2008}. In his model, the threshold change and rewiring are switched stochastically. It was shown that the adaptive thresholds yield a new class of self-organized networks. However, it was confirmed numerically that networks still evolve towards criticality in the large size limit. In summary, these previous works based on numerical simulation suggest that the activity dependent rewiring rule robustly drives networks towards criticality in different conditions. 

However, in these adaptive Boolean network models, no analytic approach has been reported so far to the best of the author's knowledge. One reason for this would be the fact that the definition of activity is dependent on attractors which are usually avoided to discuss the phase transition of Boolean networks in the limit of large system size \citep{Derrida1986a}. In the activity-dependent rewiring rule of Bornholdt and Rohlf \citep{Bornholdt2000}, a node is defined to be active if it does not change its state on the attractor reached from a random initial condition. Otherwise, the node is said to be static. The rewiring rule is as follows: The active node loses one of its incoming link randomly and the static node acquires a new incoming link randomly. Indeed, it seems that the activity on attractors is crucial for self-organization towards criticality. Bornholdt and Rohlf \citep{Bornholdt2000} numerically identified a first-order-like transition of the frozen component defined as the fraction of static nodes and argued that this transition is the main mechanism of robust self-organization of networks towards criticality. 

In this paper, we propose a new activity-dependent adaptive Boolean network model inspired by the \textit{cis}-regulatory mechanism of real-world GRNs in which activity does not dependent on attractors but is defined by typical states that will be defined in Sec.~\ref{sec:model}. By this change of the definition of activity, we expect that our model admits analysis based on a mean-field theory called the annealed approximation in the limit of large system size. In the following, we show that our model can be solved for stationary in-degree distribution for a wide class of update rules to which the annealed approximation of the Boolean dynamics is applicable. At first sight, one would suspect that our network rewiring rule is designed towards a desired result, namely, criticality. However, it turns out that whether our model can self-organize towards criticality depends on a parameter of our network rewiring rule independent of update rules. We analytically show that the average sensitivity of stationary BN dynamics is equal to the parameter if certain conditions are satisfied. Thus, only when the value of the parameter is set to $1$, we expect that BNs evolve towards criticality. The analytic result is verified by numerical simulation in four representative update rules. We also discuss the relationship between sensitivity of update rules and the tail of stationary in-degree distributions and compare it with that in real-world GRNs.

\section{Model}
\label{sec:model}
Boolean networks (BNs) consist of a directed network with $N$ nodes that can take two states $0$ and $1$. The state of node $i$ at time step $t$ is denoted by $x_i(t)$ and is updated by a rule $f_i$ selected from a given ensemble of Boolean functions $\mathcal{E}_i$: 
\begin{equation}
x_i(t+1)=f_i({\bm x}_i(t)), 
\label{eq:01}
\end{equation}
where ${\bm x}_i(t)=(x_{j_1}(t),\dots,x_{j_{k_i}}(t))$ and $j_1,\dots,j_{k_i}$ are nodes from which node $i$ receives inputs. The number of inputs $k_i$ is called in-degree of $i$. In this paper, all nodes are updated simultaneously. We also assume that the ensemble of Boolean functions $\mathcal{E}_i$ associated with node $i$ only depends on its in-degree $k_i$. 

Our activity-dependent rewiring rule for network evolution is different from those proposed in previous work \citep{Bornholdt2000, Liu2006} in the following two respects. First, both nodes and arcs can be selected at each time step of network evolution, in contrast to the previous models where only nodes are assumed to be selected. Second, we consider activity of arcs rather than that of nodes. In the previous models, activity of a selected node is measured by time-averaging its state value along a reached attractor and the decision whether the selected node gets a new incoming arc or loses an existing arc is made depending on the value of activity. In our model, when a node is selected, the node gets a new incoming arc. On the other hand, when an arc is selected, it is deleted when it is active. Here, activity of the arc is evaluated by the response of the target node $i$ to perturbations on the arc given a \textit{typical state}. That is, given an input ${\bm x}_i=(x_{j_1},\dots,x_{j_{k_i}})$ sampled randomly from a collection of states after sufficiently long time steps starting from a random initial condition, the arc is said to be active if $f_i({\bm x}_i) \neq f_i(\tilde{{\bm x}}_i)$ where $\tilde{{\bm x}}_i=(\tilde{x}_{j_1},\dots,\tilde{x}_{j_{k_i}})$ is given by $\tilde{x}_{j_l}=1-x_{j_l}$ if $j_l$ is the source of the selected arc and $\tilde{x}_{j_l}=x_{j_l}$ otherwise. These modifications are motivated by the following biological consideration: Deletion of an arc in a GRN of an organism can be caused by mutations in \textit{cis}-regulatory elements (CREs) \citep{Peter2011, Wittkopp2012} of a gene that are nearby non-coding regions of DNA where a number of proteins called transcription factors (TFs) that are themselves products of other genes can bind. TFs regulate expression of the gene by increasing or decreasing the frequency of transcription initiation. If mutations in existing CREs of a gene change the binding pattern of TFs and the expression level of the gene, it could result in undesirable behavior of the organism and the corresponding arcs in its gene regulatory network are deleted in an evolutionary time scale \citep{Rohlf2009}. On the other hand, mutations in a non-coding region of DNA within functional interaction range that is not involved in existing CREs could give rise to binding of a new TF. This means addition of a new incoming arc to the node representing the gene. Thus, nodes in a GRN can be conceived as carrying capacity to accept new incoming arcs incarnated by non-coding regions of DNA rather than coding DNA. In summary, when considering rewiring of a GRN, it is natural to treat nodes and arcs on the same footing because the physical basis of them is the same. 

In detail, our algorithm for network evolution in this paper is as follows: 
\begin{enumerate}
\item[(i)]
An initial BN with a given ensemble of Boolean functions is generated. The in-degree of each node is sampled from a Poisson distribution with mean $k_0$ and the source of each arc is chosen randomly. 
\item[(ii)]
The state of the BN is evolved from a random initial state for sufficiently long time steps to find a typical state. For any BN of finite size $N$, its state trajectory eventually falls onto an attractor. Hence, it is ideal to choose a state randomly from the attractor. However, when numerically simulating the model, it is difficult to find an attractor in a reasonable time if the BN is in the disordered phase. For efficient numerical simulation, we limit the maximum length of attractors to be detected as $T$. If no attractor is found within $2T+T'$ time steps, the last $T$ steps are stored and a state is chosen randomly from the $T$ states. In this paper, we set $T=1000$ and $T'=100$. We expect that this way of sampling a state approximates that of sampling from true typical states in the limit of large $N$ because correlations between nodes are negligible for $N \gg 1$ if the underlying network is locally tree-like and thus whether a state is on an attractor or not does not matter if it is reached after many time steps from a random initial state \citep{Drossel2008}. Indeed, this expectation accommodates to the assumptions of the mean-field theory used in Sec.~\ref{sec:analytic} and we will see that the numerically obtained in-degree distributions by this network rewiring algorithm agree well with the theoretical predictions based on the mean-field theory. 
\item[(iii)]
A particular node or a particular arc is chosen with probability $\pi_n$ or $\pi_a$, respectively. Here, we fix the ratio $\sigma:=\pi_n/\pi_a$ throughout the network evolution. If a node is chosen, then a new incoming arc is added to the node. The source of the new arc is chosen uniformly at random. If an arc is chosen, then its activity in the state chosen in step (ii) is assessed. If the arc is active, then it is deleted. Otherwise, do nothing. 
\item[(iv)]
The Boolean function on the chosen node or the target of the chosen arc in step (iii) is re-assigned following the given ensemble of Boolean functions. 
\item[(v)]
Go back to step (ii). 
\end{enumerate}
The steps (ii)-(v) constitute time unit of network evolution. We call it \textit{epoch} after \citep{Liu2006}. Note that $\pi_n N + \pi_a z(e) N=1$ should hold for all epoch $e$ where $z(e)$ is the average in-degree of the underlying directed network of BN at epoch $e$. Thus, $\pi_n=\sigma/[(\sigma+z(e))N]$ and $\pi_a=1/[(\sigma+z(e))N]$. 

In each epoch, the network topology and Boolean functions assigned are fixed as in typical applications of BNs for modeling real-world complex systems. Thus, in the above model, the time scale separation between BN dynamics and network evolution is taken for granted.

\section{Analytic results}
\label{sec:analytic}
In this section, first we develop a general mean-field theory of network evolution that can be applied to any update rule which satisfies certain conditions mentioned below. Second, we apply the analytic result derived from the mean-field theory to four update rules that have been paid attention in the literature. 

\subsection{Mean-field theory}
\label{subsec:mft}
If the large system size limit $N \to \infty$ is taken and the underlying directed network is random networks with a specified degree distribution $P(k,l)$ \citep{Newman2001}, where $P(k,l)$ is the probability that a randomly chosen node has in-degree $k$ and out-degree $l$, the stability of BN dynamics can be analyzed by a mean-field theory so-called annealed approximation \citep{Derrida1986a, Lee2007}. In the annealed approximation, correlations between nodes are neglected. This is manifested as the following ansatz taken in the mean-field calculation of BN dynamics \citep{Drossel2008}: The sources of incoming arcs to a node are chosen randomly at each time step and the Boolean functions are also re-assigned randomly at each time step. 

We apply the annealed approximation to BN dynamics in each epoch and assess its stability. For this purpose, we need to calculate sensitivity of Boolean functions selected from a given ensemble for each input \citep{Shmulevich2004}. Let $\lambda_{k,j}$ be the probability that the output of an assigned Boolean function with $k$ inputs changes when $j-$th input is flipped for $1 \leq j \leq k$. We put $\lambda_k:=\sum_{j=1}^k \lambda_{k,j}$. In general, $\lambda_{k,j}$ depends on the fraction $b_t$ of nodes with state $1$ at time step $t$. $b_t$ evolves by the following equation 
\begin{equation}
b_{t+1}=\sum_k \beta_k(b_t) P_{\rm in}(k), 
\label{eq:02}
\end{equation}
where $\beta_k(b_t)$ is the probability that the output of a node with $k$ inputs is $1$ at time step $t+1$ and $P_{\rm in}(k)=\sum_l P(k,l)$ is the in-degree distribution. Although Eq.~(\ref{eq:02}) can have periodic or chaotic solutions depending on update rules \citep{Drossel2008}, we only consider the case that Eq.~(\ref{eq:02}) has a unique stable stationary solution $b^*$ in the following. 

Now let us suppose that the dynamics of a BN settle down to the stationary regime and apply a small perturbation. Let $\tilde{d}_t$ be the fraction of damaged inputs at time step $t$. That is, $\tilde{d}_t$ is the probability that the source node of a randomly chosen arc is flipped. Neglecting the higher order terms of $\tilde{d}_t$, we obtain 
\begin{equation}
\tilde{d}_{t+1}= \lambda \tilde{d}_t 
\label{eq:03}
\end{equation}
for the time evolution of $\tilde{d}_t$ by a similar reasoning with previous work \citep{Lee2007,Squires2012}, where $\lambda=\sum_{k,l} \frac{l P(k,l)}{z} \lambda_k$ which we call \textit{average sensitivity}, $z=\sum_k k P_{\rm in}(k)$ is the average in-degree and $\lambda_k$ is evaluated at $b^*$. Let $d_t$ be the fraction of damaged nodes at time step $t$. Since $d_{t+1}=\bar{\lambda}\tilde{d}_t$ where $\bar{\lambda}=\sum_k P_{\rm in}(k)\lambda_k$, $d_t$ also follows Eq.~(\ref{eq:03}). When in-degree and out-degree are independent as we expect for networks evolved by the proposed network rewiring algorithm, we have 
\begin{equation}
\lambda=\bar{\lambda}=\sum_k P_{\rm in}(k)\lambda_k. 
\label{eq:04}
\end{equation}
When $\lambda < 1$, $d_t$ dies out eventually and the dynamics are said to be ordered or stable. If $\lambda >1 $, $d_t$ grows exponentially at first and the dynamics are said to be disordered or unstable. $\lambda=1$ is the boundary between the two cases and the dynamics are said to be critical. 

Now let us write down the equation for the time evolution of in-degree distribution by assuming the annealed approximation for the dynamics of BN at each epoch. Let $P_{\rm in}(e,k)$ be the in-degree distribution at epoch $e$. According to the proposed network rewiring algorithm, we have 
\begin{equation}
P_{\rm in}(e+1,k)= \left( 1 - \pi_n -\pi_a \lambda_k \right) P_{\rm in}(e,k) + \pi_n P_{\rm in}(e,k-1) + \pi_a \lambda_{k+1} P_{\rm in}(e,k+1) 
\label{eq:05}
\end{equation}
for $k \geq 1$ and 
\begin{equation}
P_{\rm in}(e+1,0)=\left( 1 - \pi_n \right) P_{\rm in}(e,0) + \pi_a \lambda_1 P_{\rm in}(e,1). 
\label{eq:06}
\end{equation}
In order to iteratively solve Eqs.~(\ref{eq:05}) and (\ref{eq:06}), in each iteration one must calculate $\lambda_k$ which is in general a function of $b^*$, which in turn depends on the entire in-degree distribution at epoch $e$ through Eq.~(\ref{eq:02}). In addition, $\pi_n$ and $\pi_a$ are functions of average in-degree $z(e)$. A stationary solution $P_{\rm in}^s(k)$ of Eqs.~(\ref{eq:05}) and (\ref{eq:06}) should satisfy 
\begin{equation}
\pi_n P_{\rm in}^s(k)= \pi_a \lambda_{k+1} P_{\rm in}^s(k+1) 
\label{eq:07}
\end{equation}
for $k \geq 0$ if it exists. When the stationary solution exists, we obtain 
\begin{equation}
\lambda=\pi_n/\pi_a 
\label{eq:08}
\end{equation}
by substituting Eq.~(\ref{eq:07}) into Eq.~(\ref{eq:04}). Thus, we predict that we can control the stability of evolved BNs by adjusting the ratio $\sigma=\pi_n/\pi_a$ which we call \textit{target average sensitivity} (TAS) hereafter. Note that $\sigma$ can be given independently of update rules. In particular, when $\sigma=1$, that is, when a node or an arc is selected uniformly at random, the proposed network rewiring algorithm is expected to drive BNs towards criticality. 

The limitation of our mean-field theory arises from the normalization condition for the stationary in-degree distribution. If Eq.~(\ref{eq:07}) has a solution, it is solved by 
\begin{equation}
P_{\rm in}^s(k)=P_{\rm in}^s(0) \sigma^k \left( \prod_{l=1}^k \lambda_l \right)^{-1}. 
\label{eq:09}
\end{equation}
Hence the infinite series $\sum_{k=0}^\infty r_k$ must be convergent, where $r_k=\sigma^k \left( \prod_{l=1}^k \lambda_l \right)^{-1}$. Since $r_{k+1}/r_k=\sigma/\lambda_{k+1}$, this is always the case when $\lambda_k$ diverges as $k \to \infty$ by d'Alembert's ratio test. However, when $\lambda_k$ converges to a number $\alpha$ as $k \to \infty$, it must hold that $\sigma \leq \alpha$. When $b^*$ is independent of $P_{\rm in}^s$, we can give the condition for the existence of $P_{\rm in}^s$ as follows: (i) If $\lambda_k \to \infty$ as $k \to \infty$, then $P_{\rm in}^s$ exists. (ii) If $\lambda_k \to \alpha < \infty$ as $k \to \infty$, then $P_{\rm in}^s$ exists if $\sigma<\alpha$. If $\sigma>\alpha$, then $P_{\rm in}^s$ does not exist. If $\sigma=\alpha$, then the existence of $P_{\rm in}^s$ depends on the precise form of $\lambda_k$. Even when $P_{\rm in}^s$ does not exist in the mean-field theory, we can formally obtain $P_{\rm in}^s$ by truncating Eq.~(\ref{eq:09}) at $k=N$ for BNs of finite size $N$. However, it is not guaranteed that the truncated $P_{\rm in}^s$ can reproduce the stationary in-degree distribution of the evolved finite size BNs. This is because the assumption of the absence of correlations between nodes in the annealed approximation of BN dynamics will be violated in such case due to the existence of non-negligible amount of nodes with in-degree proportional to system size $N$. 

\subsection{Examples}
\label{subsec:examples}
In this subsection, we apply the analytic result presented in Sec.~\ref{subsec:mft} to four ensembles of Boolean functions: (a) Biased functions (BF) \citep{Derrida1986a}: All Boolean functions with $k_i$ inputs are weighted with bias $p$. The value of output of $f_i$ is assigned to be $1$ with probability $p$ or $0$ with probability $1-p$ for each input ${\bm x}_i$. (b) Threshold functions (TF) \citep{Rohlf2002}: Only threshold functions are considered. $f_i({\bm x}_i)=1$ if $\sum_{l=1}^{k_i} w_{j_l i}(2x_{j_l}-1)+h_i \geq 0$ or $0$ otherwise, where ${\bm x}_i=(x_{j_1},\dots,x_{j_{k_i}}) \in \{0,1\}^{k_i}$ and $w_{j_l i}=\pm 1$ with equal probability. In the following, we only consider the case $h_i=0$ for all $i$. (c) Heterogeneous biased functions (HBF) \citep{Pomerance2009}: In this update rule, we allow the bias of BFs to depend on in-degree. That is, a BF with bias $p_{k_i}$ is selected for node $i$ with in-degree $k_i$. (d) Nested Canalizing functions (NCF) \citep{Kauffman2003}: A nested canalizing function is given by 
\begin{equation}
f({\bm x}_i)=
\begin{cases}
s_1 & \text{if $x_{j_1}=c_1$} \\
s_2 & \text{if $x_{j_1} \neq c_1$ and $x_{j_2}=c_2$} \\
s_3 & \text{if $x_{j_1} \neq c_1$ and $x_{j_2} \neq c_2$ and $x_{j_3}=c_3$} \\
\vdots & \\
s_{k_i} & \text{if $x_{j_1} \neq c_1$ and \dots and $x_{j_{k_i}}=c_{k_i}$} \\
s_d & \text{otherwise}
\end{cases}
\label{eq:10}
\end{equation}
for ${\bm x}_i=(x_{j_1},\dots,x_{j_{k_i}}) \in \{0,1\}^{k_i}$, where $c_l \in \{0,1\}$ is the canalizing value for input from node $j_l$ and $s_l \in \{0,1\}$ is the corresponding output value for $l=1,\dots,k_i$. Here, we consider a weight on NCFs defined by the following parameters \citep{Peixoto2010}: $s_l=1$ with probability $a$ and $c_l=1$ with probability $c$ for $l=1,\dots,k_i$, and $s_d=1$ with probability $d$. 

The formula of $\lambda_k$ for BFs, TFs and HBFs are given by $\lambda_k=2p(1-p)k$, $\lambda_k=k2^{-(k-1)}\binom{k-1}{\lfloor k/2 \rfloor} \sim \sqrt{2/\pi} \sqrt{k}$ \citep{Rohlf2002} and $\lambda_k=2p_k(1-p_k)k$, respectively. For these three rules, $\lambda_k$ is independent of $b^*$. However, $\lambda_k$ of NCFs depends on $b^*$. We have $\beta_k(b_t)=a+(d-a)(1-\gamma(b_t))^k$ in Eq.~(\ref{eq:02}) where $\gamma(b_t)=b_t c + (1-b_t)(1-c)$ is the probability that a randomly chosen input is at its canalizing value \citep{Peixoto2010}. $\lambda_k$ of NCFs is shown to be $\lambda_k=(1-\eta)(1-(1-\gamma(b^*))^k)/\gamma(b^*)+k(1-\gamma(b^*))^{k-1}(\eta-\eta_0) \sim (1-\eta)/\gamma(b^*)$ when $0<\gamma(b^*)<1$, where $\eta=a^2+(1-a)^2$ and $\eta_0=ad+(1-a)(1-d)$ at stationarity \citep{Peixoto2010}. 

By substituting $\lambda_k$ into the right-hand side of Eq.~(\ref{eq:09}), we obtain stationary in-degree distributions. For BFs, we get a Poisson stationary in-degree distribution $P_{\rm in}^s(k)=e^{-z_s}z_s^k/k!$ with the stationary average in-degree $z_s=\sigma/[2p(1-p)]$. The tail of the stationary in-degree distribution for TFs decays slower than that of any Poisson distribution but does faster than that of any exponential distribution. HBFs have different stationary in-degree distributions depending on the functional form of $p_k$ if it exists. For NCFs, the stationary in-degree distribution exists and is asymptotically equal to an exponential distribution provided that $0<\gamma(b^*)<1$ and $\sigma < (1-\eta)/\gamma(b^*)$ where $b^*$ satisfies $b^*=\sum_k \beta_k(b^*)P_{\rm in}^s(k)$. 

In next section, we test these analytic predictions for TAS $\sigma$ close to $1$ since our primary interest is evolution towards criticality. The behavior of our model for a wider range of $\sigma$ is investigated in Appendix where we also present an example in which our mean-field theory fails. 

\begin{figure}[t]
\begin{center}
\includegraphics[width=120mm]{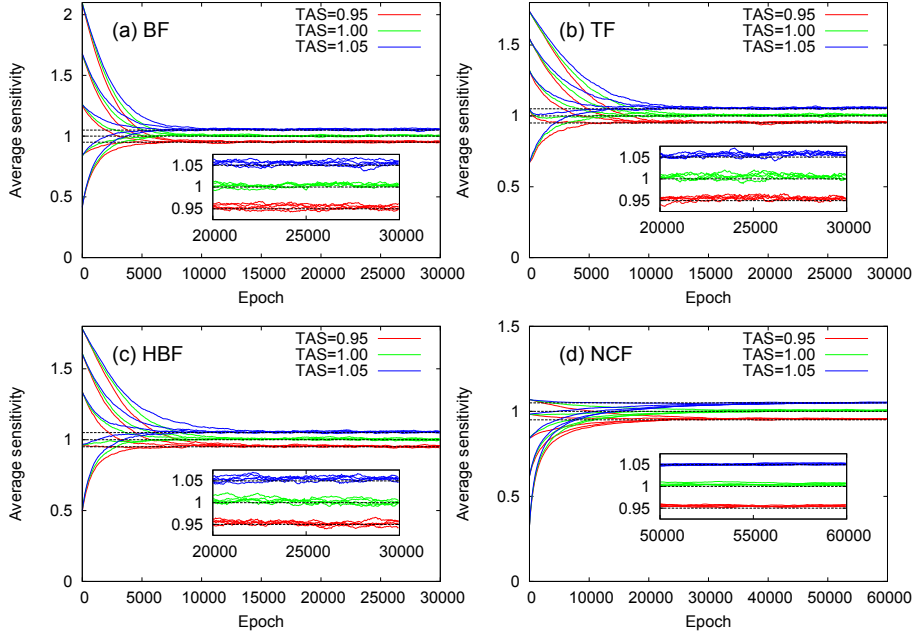}
\caption{
Time evolution of the average sensitivities for (a) BFs, (b) TFs, (c) HBFs and (d) NCFs. Insets are enlarged views from epoch $20000$ to $30000$ for the first three update rules and that from $50000$ to $60000$ for NCFs. BNs with NCFs were simulated for a longer period because their convergence is slower than the others. 
\label{fig1}}
\end{center}
\end{figure}

\begin{figure}[t]
\begin{center}
\includegraphics[width=120mm]{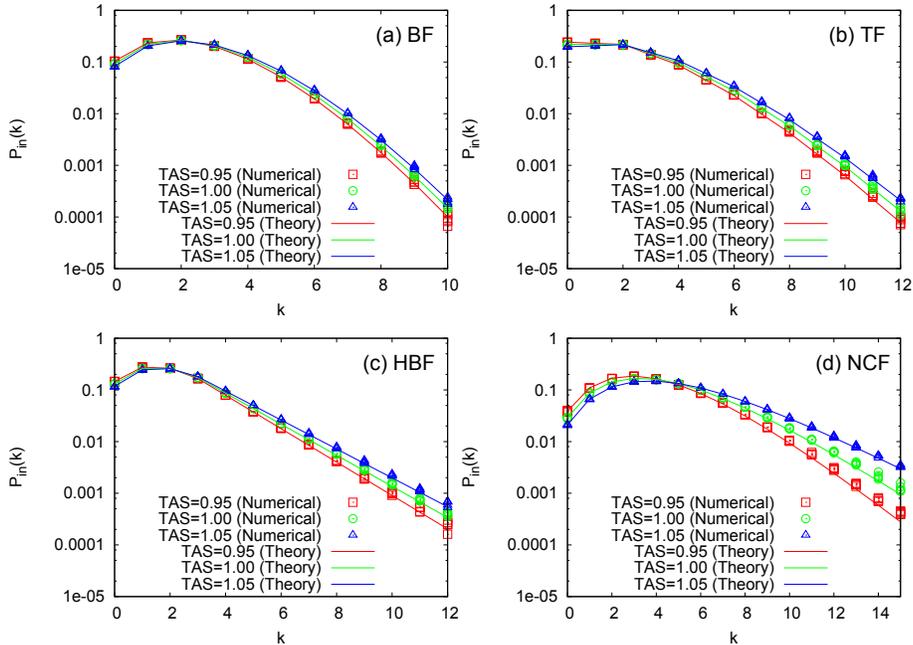}
\caption{
Comparison between numerical stationary in-degree distributions (symbols) and theoretical stationary in-degree distributions (lines) for (a) BFs, (b) TFs, (c) HBFs and (d) NCFs. 
\label{fig2}}
\end{center}
\end{figure}

\begin{figure}[t]
\begin{center}
\includegraphics[width=120mm]{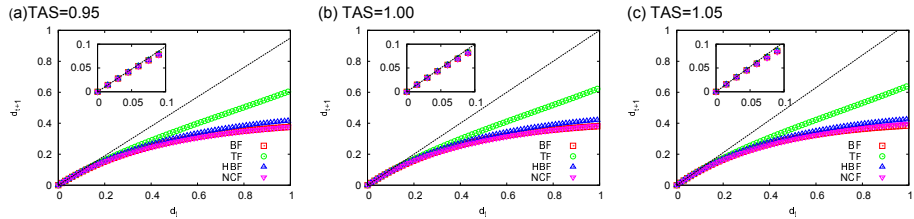}
\caption{
The fraction of damaged nodes $d_{t+1}$ at time step $t+1$ as a function of the fraction of damaged nodes $d_t$ at time step $t$ for (a) $\sigma=0.95$, (b) $\sigma=1.00$ and (c) $\sigma=1.05$. Dotted lines have a slope equal to $\sigma$. Insets are enlarged views of the region $0 \leq d_t, d_{t+1} \leq 0.1$. 
\label{fig3}}
\end{center}
\end{figure}

\section{Numerical results}
\label{sec:numerical}
We compared analytic results with numerical simulations for the above four ensembles of Boolean functions. We simulated evolution of BNs with $N=200$ for three different values of TAS: $\sigma=0.95, 1.00$ and $1.05$. Parameters used are $p=0.7$ for BFs, $p_k=(1+\sqrt{1-2q_k})/2$ with $q_k=1/2$ if $1 \leq k \leq 3$ and $q_k=2/k$ if $k \geq 4$ for HBFs (thus, we have $\lambda_k=2$ for $k \geq 4$) and $a=1/3$, $c=0.95$ and $d=0$ for NCFs. The condition for the existence of the stationary in-degree distribution for HBFs is $\sigma < 2$ and is satisfied in the numerical simulation here. For NCFs, we numerically checked that $0<\gamma(b^*)<1$ and $\sigma < (1-\eta)/\gamma(b^*)$ hold for the above parameter values. 

Fig.~\ref{fig1} shows time evolution of the average sensitivities for each update rule from five different initial average in-degree $1 \leq k_0 \leq 5$. For each pair of values of $\sigma$ and $k_0$, $100$ realizations were averaged. In Fig.~\ref{fig1}, the average sensitivity of a BN at epoch $e$ was calculated by Eq.~(\ref{eq:04}) with a numerical in-degree distribution at epoch $e$ and analytic values of $\lambda_k$. We can clearly see that the average sensitivities approach to given values of $\sigma$ independent of $k_0$. 

The numerical stationary in-degree distributions agree well with the theoretical predictions (Eq.~(\ref{eq:09})) for all three values of TAS $\sigma$ (Fig.~\ref{fig2}). Here, they were obtained by averaging numerical in-degree distributions over last $10000$ epochs in Fig.~\ref{fig1} of $100$ realizations for each $k_0$. 

Finally, we verified numerically that Eq.~(\ref{eq:03}) (with replacing $\tilde{d}_{t}$ and $\tilde{d}_{t+1}$ by $d_t$ and $d_{t+1}$, respectively) holds in evolved BNs for all three values of TAS $\sigma$ by constructing so-called Derrida plots (Fig.~\ref{fig3}) \citep{Derrida1986b}. Derrida plots show the fraction of damaged nodes $d_{t+1}$ at time step $t+1$ as a function of the fraction of damaged nodes $d_t$ at time step $t$. In Fig.~\ref{fig3}, the value of $d_{t+1}$ was averaged over $200$ states of $500$ realizations of evolved BNs (those at the last step in Fig.~\ref{fig1}) for each value of $d_t$. We can see that for all three values of TAS, the slope at the origin agrees well between numerical calculations and theoretical predictions. In constructing Derrida plots numerically, a subtlety arises when $\lambda_k$ depends on $b^*$ as in case of NCFs. For BFs, TFs and HBFs, we can choose a random state and randomly flip its fraction of $d_t$ nodes to compute $d_{t+1}$ because $\lambda_k$ is independent of $b^*$ in these update rules. On the other hand, for NCFs, we must choose a typical state and then randomly flip its fraction of $d_t$ nodes. It was predicted that this procedure produces the correct slope at the origin of Derrida plots \citep{Kesseli2006}. However, in order for a Derrida plot to be correct for larger values of $d_t$, the perturbed state must also be a random sample of typical states (This does not guarantee that the Derrida plot is correct over all the range of $d_t$ as shown in \citep{Kesseli2006}). Here, we are interested in only the slope of the Derrida plots at the origin. Hence, it suffices for our purpose to adopt the above procedure. 

\section{Discussion}
\label{sec:discuss}
In this paper, we proposed a new activity-dependent adaptive Boolean network model and presented its analytic solutions for stationary in-degree distribution by employing the annealed approximation of Boolean dynamics. We showed analytically that stationary BNs evolved by the proposed network rewiring algorithm have in-degree distributions whose average sensitivity is equal to TAS if certain conditions are satisfied and verified the analytic solutions agree well with numerical simulations for four representative update rules. We emphasize that TAS can be given independently of update rules. In particular, if it is set to $1$, our mean-field theory predicts that BNs evolve towards criticality. 

In previous work \citep{Hesse2014,Markovic2014}, network self-organization towards criticality has been explained by the self-organized criticality picture \citep{Bak1987,Jensen1998}. That is, criticality is achieved by slowly adding links in the subcritical phase and rapidly deleting links in the supercritical phase of an absorbing transition of network activity. In particular, Droste et al. \citep{Droste2013} analytically demonstrated this mechanism based on the pair-approximation of the network activity dynamics. They showed that two different time-scale separations are necessary to realize self-organization towards criticality: one is that between state dynamics on networks and topological changes of networks and the other is that between deletion of links and addition of links. In our model, the former time-scale separation is incorporated. However, the latter does not hold because the ratio of the probability of link addition to that of link deletion is finite. Thus, the self-organized criticality picture seems not to hold. In our model, the criticality is realized by stochastically balancing the mutually opposed processes, addition and deletion of links. 

In previous work on activity-dependent adaptive Boolean networks, influence of the update rule on the structure of evolved networks is assessed by only numerical simulations \citep{Liu2006,Rohlf2009}. In our model, we have a simple relationship between the sensitivity of update rules represented by $\lambda_k$ and the stationary in-degree distribution as shown above. Although our model is parsimonious, it is worth to compare our result with real-world GRNs. The in-degree distribution of the prokaryote \textit{Escherichia coli} is best fitted by a Poisson distribution, whereas that of the eukaryote \textit{Saccharomyces cerevisiae} is best fitted by an exponential distribution \citep{Aldana2007}. As for update rules, NCFs were introduced to model the yeast GRN \citep{Kauffman2003} because NCFs are found abundantly in eukaryotic GRNs by an extensive literature study \citep{Harris2002}. On the other hand, the analysis by Balleza et al.~\citep{Balleza2008} suggested that BFs are enough to model the GRN of \textit{E. coli}. They modeled several real-world GRNs including the bacterium GRN by biased functions to reveal whether they operate close to criticality or not and showed that changes in the fraction of canalizing functions for genes with at least 4 inputs do not affect the near critical dynamical behavior of the bacterium GRN. On the other hand, most of genes in the bacterium GRN have at most 3 inputs and canalizing functions are abundant just by chance for such genes \citep{Balleza2008}. Thus, there is no need for the bacterium to bias the sampling strategy of update rules towards canalizing functions even if they have an evolutionary advantage. Our model predicts Poisson and exponential stationary in-degree distributions for BFs and NCFs, respectively, and thus is consistent with the real-world GRNs. 

We are almost ignorant of out-degree distributions in this paper. Under the proposed network rewiring algorithm, the stationary out-degree distribution becomes a Poisson distribution independent of update rules. This disagrees with real-world GRNs because they have heavy-tailed out-degree distributions \citep{Aldana2007}. However, we can control the shape of stationary out-degree distribution by modifying step (iii) of the algorithm without changing the value of average sensitivity: selecting the source of a new arc following an appropriate weight depending on the out-degree of each node \citep{Haruna2014}. 

Finally, we note that it is an interesting open question whether our model can be extended to the network ensembles to which the semi-annealed approximation of Boolean dynamics \citep{Pomerance2009, Squires2014} is applicable.

\section*{Acknowledgments}
This work was partially supported by JSPS KAKENHI Grant Number 25280091. The author thanks the anonymous reviewers for their helpful comments to improve the manuscript.

\appendix
\section*{Appendix}
\label{sec:append}
In this appendix, we compare our theoretical results with numerical simulation for BFs and HBFs for TAS $\sigma$ apart from criticality. The parameters of the update rules are the same as those in Sec.~\ref{sec:numerical}. Our theory predicts that $P_{\rm in}^s$ exists for any $\sigma$ for BFs, while exists only for $\sigma < 2$ for HBFs since $\lambda_k=2$ for large $k$. The condition of numerical simulation is the same as that in Sec.~\ref{sec:numerical} except that we only show results for $k_0=3$ here. 

\begin{figure}[t]
\begin{center}
\includegraphics[width=120mm]{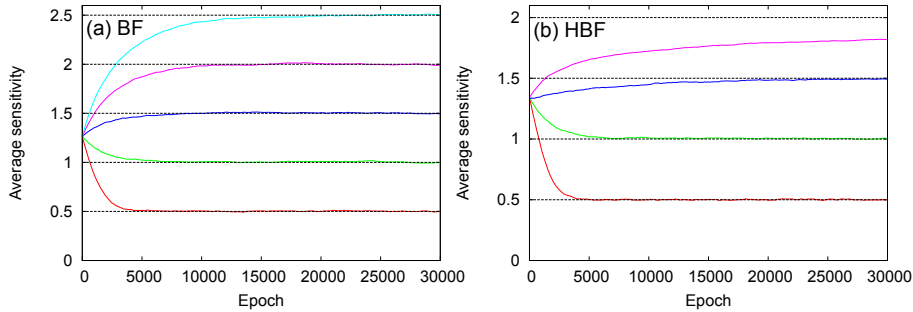}
\caption{
Time evolution of the average sensitivity. (a) BFs. The values of TAS $\sigma$ are $0.5$, $1.0$, $1.5$, $2.0$ and $2.5$ from below. (b) HBFs. The values of TAS $\sigma$ are $0.5$, $1.0$, $1.5$ and $2.0$ from below. 
\label{fig_a1}}
\end{center}
\end{figure}

\begin{figure}[htbp]
\begin{center}
\includegraphics[width=120mm]{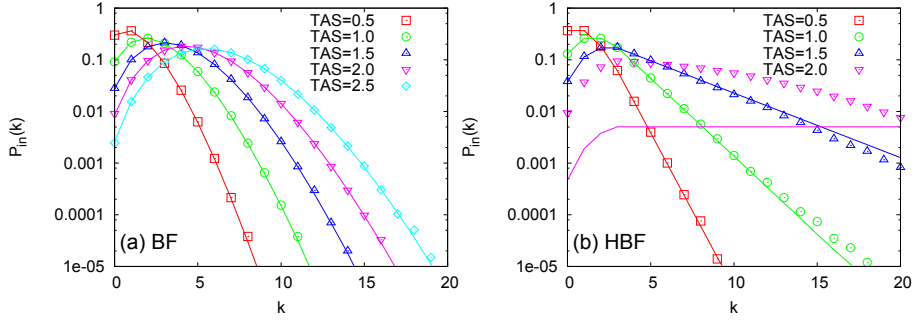}
\caption{
Comparison between numerical stationary in-degree distributions (symbols) and theoretical stationary in-degree distributions (lines) for (a) BFs and (b) HBFs. The theoretical stationary in-degree distribution does not exist for HBFs with $\sigma=2.0$. The shown theoretical line is calculated by Eq.~(\ref{eq:09}) and is truncated by $N=200$. 
\label{fig_a2}}
\end{center}
\end{figure}

\begin{figure}[htbp]
\begin{center}
\includegraphics[width=120mm]{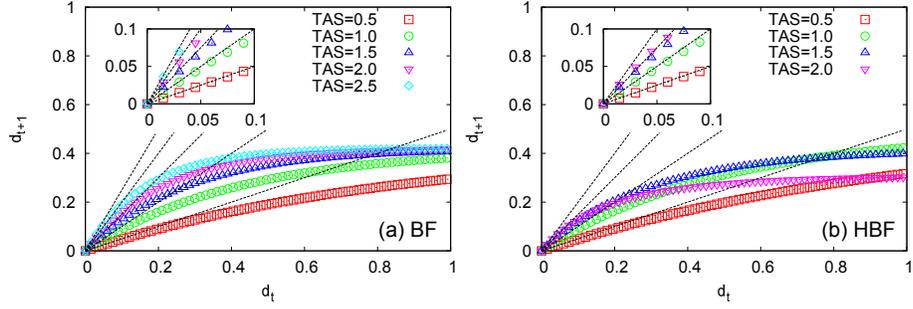}
\caption{
The fraction of damaged nodes $d_{t+1}$ at time step $t+1$ as a function of the fraction of damaged nodes $d_t$ at time step $t$ for (a) BFs and (b) HBFs. Dotted lines have the same slope as corresponding value of TAS $\sigma$. Insets are enlarged views of the region $0 \leq d_t, d_{t+1} \leq 0.1$. 
\label{fig_a3}}
\end{center}
\end{figure}

In Fig.~\ref{fig_a1}, time evolution of the average sensitivity for BFs and HBFs is shown. We can see that the average sensitivity approaches to each specified value of $\sigma$ except $\sigma=2.0$ for HBFs. The failure of evolution of BNs with HBFs towards TAS $\sigma=2.0$ can also be seen from its in-degree distribution (Fig.~{\ref{fig_a2}}) and the Derrida plot (Fig.~{\ref{fig_a3}}). On the other hand, theoretical predictions and results of numerical simulation for the other cases agree well in both Fig.~{\ref{fig_a2}} and Fig.~{\ref{fig_a3}}. These results provide further support for the claim at the end of Sec.~\ref{subsec:mft} on the applicability and the limitation of our mean-field theory. 

\newpage


\begin{thebibliography}{51}
\providecommand{\natexlab}[1]{#1}
\providecommand{\url}[1]{\texttt{#1}}
\expandafter\ifx\csname urlstyle\endcsname\relax
  \providecommand{\doi}[1]{doi: #1}\else
  \providecommand{\doi}{doi: \begingroup \urlstyle{rm}\Url}\fi

\bibitem[Drossel(2008)]{Drossel2008}
B.~Drossel.
\newblock Random {B}oolean networks.
\newblock In H.~G. Schuster, editor, \emph{Reviews of Nonlinear Dynamics and
  Complexity}. Wiley-VCH Verlag GmbH \& Co. KGaA, Weinheim, Germany, 2008.

\bibitem[Kauffman(1969)]{Kauffman1969}
S.~A. Kauffman.
\newblock Metablic stability and epigenesis in randomly constructed genetic
  nets.
\newblock \emph{J. Theor. Biol.}, 22:\penalty0 437--467, 1969.

\bibitem[K\"{u}rten(1988)]{Kurten1988}
K.~E. K\"{u}rten.
\newblock Critical phenomena in model neural networks.
\newblock \emph{Phys. Lett. A}, 129:\penalty0 157--160, 1988.

\bibitem[Paczuski et~al.(2000)Paczuski, Bassler, and Corral]{Paczuski2000}
M.~Paczuski, K.~E. Bassler, and \'{A} Corral.
\newblock Self-organized networks of competing boolean agents.
\newblock \emph{Phys. Rev. Lett.}, 84:\penalty0 3185--3188, 2000.

\bibitem[Derrida and Pomeau(1986)]{Derrida1986a}
B.~Derrida and Y.~Pomeau.
\newblock Random networks of automata: A simple annealed approximation.
\newblock \emph{Europhys. Lett.}, 1:\penalty0 45--49, 1986.

\bibitem[Beggs and Plenz(2003)]{Beggs2003}
J.~M. Beggs and D.~Plenz.
\newblock Neuronal avalanches in neocortical circuits.
\newblock \emph{J. Neurosci.}, 23:\penalty0 11167--11177, 2003.

\bibitem[Petermann et~al.(2009)Petermann, Thiagarajan, Lebedev, Nicolelis,
  Chialvo, and Plenz]{Petermann2009}
T.~Petermann, T.~C. Thiagarajan, M.~A. Lebedev, M.~A.~L. Nicolelis, D.~R.
  Chialvo, and D.~Plenz.
\newblock Spontaneous cortical activity in awake monkeys composed of neuronal
  avalanches.
\newblock \emph{Proc. Natl. Acad. Sci. U.S.A.}, 106:\penalty0 15921--15926,
  2009.

\bibitem[Balleza et~al.(2008)Balleza, Alvarez-Buylla, Chaos, Kauffman,
  Shmulevich, and Aldana]{Balleza2008}
E.~Balleza, E.~R. Alvarez-Buylla, A.~Chaos, S.~Kauffman, I.~Shmulevich, and
  M.~Aldana.
\newblock Critical dynamics in genetic regulatory networks: Examples from four
  kingdoms.
\newblock \emph{PLoS ONE}, 3:\penalty0 e2456, 2008.

\bibitem[Nykter et~al.(2008)Nykter, Price, Aldana, Ramsey, Kauffman, Hood,
  Yli-Harja, and Shmulevich]{Nykter2008}
M.~Nykter, N.D. Price, M.~Aldana, S.~A. Ramsey, S.~A. Kauffman, L.~E. Hood,
  O.~Yli-Harja, and I.~Shmulevich.
\newblock Gene expression dynamics in the macrophage exhibit criticality.
\newblock \emph{Proc. Natl. Acad. Sci. U.S.A.}, 105:\penalty0 1897--1900, 2008.

\bibitem[Valverde et~al.(2015)Valverde, Ohse, Turalska, Garcia-Ojalvo, and
  West]{Valverde2015}
S.~Valverde, S.~Ohse, M.~Turalska, J.~Garcia-Ojalvo, and B.~J. West.
\newblock Structural determinants of criticality in biological networks.
\newblock \emph{Front. Physiol.}, 6:\penalty0 127, 2015.

\bibitem[Bertschinger and Natschl\"{a}ger(2004)]{Bertschinger2004}
N.~Bertschinger and T.~Natschl\"{a}ger.
\newblock Real-time computation at the edge of chaos in recurrent neural
  networks.
\newblock \emph{Neural Comput.}, 16:\penalty0 1413--1436, 2004.

\bibitem[Goudarzi et~al.(2012)Goudarzi, Teuscher, Gulbahce, and
  Rohlf]{Goudarzi2012}
A.~Goudarzi, C.~Teuscher, N.~Gulbahce, and T.~Rohlf.
\newblock Emergent criticality through adaptive information processing in
  boolean networks.
\newblock \emph{Phys. Rev. Lett.}, 108:\penalty0 128702, 2012.

\bibitem[Kinouchi and Copelli(2006)]{Kinouchi2006}
O.~Kinouchi and A.~M. Copelli.
\newblock Optimal dynamical range of excitable networks at criticality.
\newblock \emph{Nature Physics}, 2:\penalty0 348--352, 2006.

\bibitem[Haldeman and Beggs(2005)]{Haldeman2005}
C.~Haldeman and J.~M. Beggs.
\newblock Critical branching captures activity in living neural networks and
  maximizes the number of metastable states.
\newblock \emph{Phys. Rev. Lett.}, 94:\penalty0 058101, 2005.

\bibitem[Bornholdt and R\"{o}hl(2003)]{Bornholdt2003}
S.~Bornholdt and T.~R\"{o}hl.
\newblock Self-organized critical neural networks.
\newblock \emph{Phys. Rev. E}, 67:\penalty0 066118, 2003.

\bibitem[Rybarsch and Bornholdt(2014)]{Rybarsch2014}
M.~Rybarsch and S.~Bornholdt.
\newblock Avalanches in self-organized critical neural networks: A minimal
  model for the neural soc universality class.
\newblock \emph{PLoS ONE}, 9:\penalty0 e93090, 2014.

\bibitem[Meisel and Gross(2009)]{Meisel2009}
C.~Meisel and T.~Gross.
\newblock Adaptive self-organization in a realistic neural network model.
\newblock \emph{Phys. Rev. E}, 80:\penalty0 061917, 2009.

\bibitem[Rubinov et~al.(2011)Rubinov, Sporns, Thivierge, and
  Breakspear]{Rubinov2011}
M.~Rubinov, O.~Sporns, J.-P. Thivierge, and M.~Breakspear.
\newblock Neurobiologically realistic determinants of self-organized
  criticality in networks of spiking neurons.
\newblock \emph{PLoS. Comput. Biol.}, 7:\penalty0 e1002038, 2011.

\bibitem[Levina et~al.(2007)Levina, Herrmann, and Geisel]{Levina2007}
A.~Levina, J.~M. Herrmann, and T.~Geisel.
\newblock Dynamical synapses causing self-organized criticality in neural
  networks.
\newblock \emph{Nature Phys.}, 3:\penalty0 857--860, 2007.

\bibitem[Levina et~al.(2009)Levina, Herrmann, and Geisel]{Levina2009}
A.~Levina, J.~M. Herrmann, and T.~Geisel.
\newblock Phase transitions towards criticality in a neural system with
  adaptive interactions.
\newblock \emph{Phys. Rev. Lett.}, 102:\penalty0 118110, 2009.

\bibitem[Droste et~al.(2013)Droste, Do, and Gross]{Droste2013}
F.~Droste, A.-L. Do, and T.~Gross.
\newblock Analytical investigation of self-organized criticality in neural
  networks.
\newblock \emph{J. R. Soc. Interface}, 10:\penalty0 20120558, 2013.

\bibitem[MacArthur et~al.(2010)MacArthur, S\'{a}nchez-Garc\'{i}a, and
  Ma'ayan]{MacArthur2010}
B.~D. MacArthur, R.~J. S\'{a}nchez-Garc\'{i}a, and A.~Ma'ayan.
\newblock Microdynamics and criticality of adaptive regulatory networks.
\newblock \emph{Phys. Rev. Lett.}, 104:\penalty0 168701, 2010.

\bibitem[Lee(2014)]{Lee2014}
D-S. Lee.
\newblock Evolution of regulatory networks towards adaptability and stability
  in a changing environment.
\newblock \emph{Phys. Rev. E}, 90:\penalty0 052822, 2014.

\bibitem[Gross and Sayama(2009)]{Gross2009}
T.~Gross and H.~Sayama, editors.
\newblock \emph{Adaptive Networks: Theory,Models and Applications}.
\newblock Springer Verlag, Heidelberg, 2009.

\bibitem[Sayama et~al.(2013)Sayama, Pestov, Schmidt, Bush, Wong, Yamanoi, and
  Gross]{Sayama2013}
H.~Sayama, I.~Pestov, J.~Schmidt, B.~J. Bush, C.~Wong, J.~Yamanoi, and
  T.~Gross.
\newblock Modeling complex systems with adaptive networks.
\newblock \emph{Comput. Math. Appl.}, 65:\penalty0 1645--1664, 2013.

\bibitem[Bornholdt and Rohlf(2000)]{Bornholdt2000}
S.~Bornholdt and T.~Rohlf.
\newblock Topological evolution of dynamical networks: Global criticality from
  local dynamics.
\newblock \emph{Phys. Rev. Lett.}, 84:\penalty0 6114--6117, 2000.

\bibitem[Christensen et~al.(1998)Christensen, Donangelo, Koiller, and
  Sneppen]{Christensen1998}
K.~Christensen, R.~Donangelo, B.~Koiller, and K.~Sneppen.
\newblock Evolution of random networks.
\newblock \emph{Phys. Rev. Lett.}, 81:\penalty0 2380--2383, 1998.

\bibitem[Liu and Bassler(2006)]{Liu2006}
M.~Liu and K.~E. Bassler.
\newblock Emergent criticality from coevolution in random boolean networks.
\newblock \emph{Phys. Rev. E}, 74:\penalty0 041910, 2006.

\bibitem[G\'{o}rski et~al.(2016)G\'{o}rski, Czaplicka, and
  Ho{\l}yst]{Gorski2016}
P.~J. G\'{o}rski, A.~Czaplicka, and A.~Ho{\l}yst.
\newblock Coevolution of information processing and topology in hierarchical
  adaptive random boolean networks.
\newblock \emph{Eur. Phys. J. B}, 89:\penalty0 33, 2016.

\bibitem[Rohlf(2008)]{Rohlf2008}
T.~Rohlf.
\newblock Self-organization of heterogeneous topology and symmetry breaking in
  networks with adaptive thresholds and rewiring.
\newblock \emph{Europhys. Lett.}, 84:\penalty0 10004, 2008.

\bibitem[Peter and Davidson(2011)]{Peter2011}
I.~S. Peter and E.~H. Davidson.
\newblock Evolution of gene regulatory networks controlling body plan
  development.
\newblock \emph{Cell}, 144:\penalty0 970--985, 2011.

\bibitem[Wittkopp and Kalay(2012)]{Wittkopp2012}
P.~J. Wittkopp and G.~Kalay.
\newblock Cis-regulatory elements: molecular mechanisms and evolutionary
  processes underlying divergence.
\newblock \emph{Nat. Rev. Genet.}, 13:\penalty0 59--69, 2012.

\bibitem[Rohlf and Bornholdt(2009)]{Rohlf2009}
T.~Rohlf and S.~Bornholdt.
\newblock Self-organized criticality and adaptation in discrete dynamical
  networks.
\newblock In T.~Gross and H.~Sayama, editors, \emph{Adaptive Networks}, pages
  73--106. Springer, Heidelberg, Germany, 2009.

\bibitem[Newman et~al.(2001)Newman, Strogatz, and Watts]{Newman2001}
M.~E.~J. Newman, S.~H. Strogatz, and D.~J. Watts.
\newblock Random graphs with arbitrary degree distributions and their
  applications.
\newblock \emph{Phys. Rev. E}, 64:\penalty0 026118, 2001.

\bibitem[Lee and Rieger(2007)]{Lee2007}
D.-S. Lee and H.~Rieger.
\newblock Comparative study of the transcriptional regulatory networks of e.
  coli and yeast: Structural characteristics leading to marginal dynamic
  stability.
\newblock \emph{J. Theor. Biol.}, 248:\penalty0 618--626, 2007.

\bibitem[Shmulevich and Kauffman(2004)]{Shmulevich2004}
I.~Shmulevich and S.~A. Kauffman.
\newblock Activities and sensitivities in {B}oolean network models.
\newblock \emph{Phys. Rev. Lett.}, 93:\penalty0 048701, 2004.

\bibitem[Squires et~al.(2012)Squires, Ott, and Girvan]{Squires2012}
S.~Squires, E.~Ott, and M.~Girvan.
\newblock Dynamical instability in {B}oolean networks as a percolation problem.
\newblock \emph{Phys. Rev. Lett.}, 109:\penalty0 085701, 2012.

\bibitem[Rohlf and Bornholdt(2002)]{Rohlf2002}
T.~Rohlf and S.~Bornholdt.
\newblock Criticality in random threshold networks: annealed approximation and
  beyond.
\newblock \emph{Physica A}, 310:\penalty0 245--259, 2002.

\bibitem[Pomerance et~al.(2009)Pomerance, Ott, Girvan, and
  Losert]{Pomerance2009}
A.~Pomerance, E.~Ott, M.~Girvan, and W.~Losert.
\newblock The effect of network topology on the stability of discrete state
  models of genetic control.
\newblock \emph{Proc. Natl. Acad. Sci. USA}, 106:\penalty0 8209--8214, 2009.

\bibitem[Kauffman et~al.(2003)Kauffman, Peterson, Samuelsson, and
  Troein]{Kauffman2003}
S.~Kauffman, C.~Peterson, B.~Samuelsson, and C.~Troein.
\newblock Random {B}oolean network models and the yeast transcriptional
  network.
\newblock \emph{Proc. Natl. Acad. Sci. USA}, 100:\penalty0 14796--14799, 2003.

\bibitem[Peixoto(2010)]{Peixoto2010}
T.~P. Peixoto.
\newblock The phase diagram of random {B}oolean networks with nested canalizing
  functions.
\newblock \emph{Eur. Phys. J. B}, 78:\penalty0 187--192, 2010.

\bibitem[Derrida and Weisbuch(1986)]{Derrida1986b}
B.~Derrida and G.~Weisbuch.
\newblock Evolution of overlaps between configurations in random {B}oolean
  networks.
\newblock \emph{J. Phys.}, 47:\penalty0 1297--1303, 1986.

\bibitem[Kesseli et~al.(2006)Kesseli, R\"{a}m\"{o}, and Yli-Harja]{Kesseli2006}
J.~Kesseli, P.~R\"{a}m\"{o}, and O.~Yli-Harja.
\newblock Iterated maps for annealed {B}oolean networks.
\newblock \emph{Phys. Rev. E}, 74:\penalty0 046104, 2006.

\bibitem[Hesse and Gross(2014)]{Hesse2014}
J.~Hesse and T.~Gross.
\newblock Self-organized criticality as a fundamental property of neural
  systems.
\newblock \emph{Front. Syst. Neurosci.}, 8:\penalty0 1--14, 2014.

\bibitem[Markovi\'{c} and Gros(2014)]{Markovic2014}
D.~Markovi\'{c} and C.~Gros.
\newblock Power laws and self-organized criticality in theory and nature.
\newblock \emph{Phys. Rep.}, 536:\penalty0 41--74, 2014.

\bibitem[Bak et~al.(1987)Bak, Tang, and Wiesenfeld]{Bak1987}
P.~Bak, C.~Tang, and K.~Wiesenfeld.
\newblock Self-organized criticality: an explanation of $1/f$ noise.
\newblock \emph{Phys. Rev. Lett.}, 59:\penalty0 381--384, 1987.

\bibitem[Jensen(1998)]{Jensen1998}
H.~J. Jensen.
\newblock \emph{Self-Organized Criticality: Emergent Complex Behavior in
  Physical and Biological Systems}.
\newblock Cambridge Univ. Press, Cambridge, 1998.

\bibitem[Aldana et~al.(2007)Aldana, Balleza, Kauffman, and
  Resendiz]{Aldana2007}
M.~Aldana, E.~Balleza, S.~Kauffman, and O.~Resendiz.
\newblock Robustness and evolvability in genetic regulatory networks.
\newblock \emph{J. Theor. Biol.}, 245:\penalty0 433--448, 2007.

\bibitem[Harris et~al.(2002)Harris, Sawhill, Wuensche, and
  Kauffman]{Harris2002}
S.~E. Harris, B.~K. Sawhill, A.~Wuensche, and S.~Kauffman.
\newblock A model of transcriptional regulatory networks based on biases in the
  observed regulation rules.
\newblock \emph{Complexity}, 7:\penalty0 23--40, 2002.

\bibitem[Haruna and Tanaka(2014)]{Haruna2014}
T.~Haruna and S.~Tanaka.
\newblock On the relationship between local rewiring rules and stationary
  out-degree distributions in adaptive random boolean network models.
\newblock In H.~Sayama, J.~Rieffel, S.~Risi, R.~Doursat, and H.~Lipson,
  editors, \emph{Artificial Life 14: Proceedings of the Fourteenth
  International Conference on the Synthesis and Simulation of Living Systems},
  pages 419--426. MIT Press, Cambridge, 2014.

\bibitem[Squires et~al.(2014)Squires, Pomerance, Girvan, and Ott]{Squires2014}
S.~Squires, A.~Pomerance, M.~Girvan, and E.~Ott.
\newblock Stability of {B}oolean networks: The joint effects of topology and
  update rules.
\newblock \emph{Phys. Rev. E}, 90:\penalty0 022814, 2014.

\end{thebibliography}

\end{document}